\documentclass[iop,apjl]{emulateapj}

\usepackage{epstopdf}

\newcommand{\sgr}{\mbox{Sgr\ A$^\ast$}}

\usepackage{natbib}

\slugcomment{To appear in ApJL: submitted 2014 Sep 1, accepted 2014 Sep 4}

\shorttitle{\sgr\ Substructure}
\shortauthors{Gwinn et al.}

\begin{document}

\title{Discovery of Substructure in the \\ Scatter-Broadened Image of Sgr A*}

\author{C.R. Gwinn\altaffilmark{1}, Y.Y. Kovalev\altaffilmark{2,3}, M.D. Johnson\altaffilmark{4}, and V.A. Soglasnov\altaffilmark{2}}

\altaffiltext{1}{Physics Department, Broida Hall, University of California, Santa Barbara California 93117, USA}
\altaffiltext{2}{Astro Space Center, Lebedev Physical Institute, Russian Academy of Sciences, Profsoyuznaya str. 84/32, Moscow 117997, Russia}
\altaffiltext{3}{Max Planck Institute for Radio Astronomy, Auf dem H\"ugel 69, D-53121 Bonn, Germany}
\altaffiltext{4}{Harvard-Smithsonian Center for Astrophysics, 60 Garden St, Cambridge MA 02138, USA}

\begin{abstract}
We have detected substructure within the smooth scattering disk of the celebrated Galactic Center radio source Sagittarius A* (\sgr).  
We observed this structure at 1.3 cm wavelength with the Very Long Baseline Array together with the Green Bank Telescope, 
on baselines of up to 3000 km, long enough to completely resolve the average scattering disk. 
Such structure is predicted theoretically, as a consequence of refraction by large-scale plasma fluctuations in the interstellar medium. 
Along with the much-studied $\theta_{\rm d}\propto \lambda^2$ scaling of angular broadening $\theta_{\rm d}$ with observing wavelength $\lambda$, 
our observations indicate that the spectrum of interstellar turbulence is shallow, with an inner scale larger than 300\ km.
The substructure is consistent with an intrinsic size of about 1\ mas  at 1.3 cm wavelength, as inferred from deconvolution of the average scattering.
Further observations of the substructure can set stronger constraints on the properties of scattering material and on the intrinsic size of \sgr.
These constraints will guide understanding of effects of scatter-broadening and emission physics 
of the black hole, 
in images with the Event Horizon Telescope at millimeter wavelengths.\end{abstract}


\section{Introduction}

\subsection{\sgr}

\sgr\ marks a super-massive black hole, of mass $4.5\times 10^6\ M_{\odot}$ 
in the center of the Milky Way at a distance of 8.4 kpc \citep{Reid09,Ghez08}.
Its close distance and wide spectral range of emission make it an excellent subject for studies to understand the supermassive black holes believed to lie at the core of every galaxy \citep{Ric98,Ho08}.
\sgr\ shows emission at radio, infrared and X-ray wavelengths similar to that of the dramatic active nuclei of other galaxies, but with much lower luminosity
 \citep{Fal98,Gen03,Bag03,Ghez04}. \sgr\ appears to be in a relatively quiescent state, raising interesting issues of the origin of its emission and its coupling with the surrounding matter.
The spectrum, size and variability are consistent with accretion onto a supermassive black hole \citep{Yuan02}. 

Current models of the radio emission typically invoke an inefficient accretion flow \citep{Nar95,Nar98}, a jet \citep{Fal00,Mar07}, or a composite \citep{Yuan02,Mos13}. Pure accretion flow models significantly underpredict the centimeter-wavelength flux from Sgr A* without the addition of a small non-thermal electron population \citep{Maha98,Ozel03,Yuan03}. For all models, the intrinsic size increases with wavelength, reflecting the changing location of the photosphere. However, details of the emission morphology are strikingly different -- a jet feature will be highly anisotropic, while emission from a non-thermal population may exhibit a limb-brightened shell. The intrinsic structure of \sgr\ at centimeter to millimeter wavelengths characterizes regions of the source that are weak or invisible at shorter wavelengths. Hence, measurements of intrinsic size over a broad range of wavelengths are essential for assembling a global picture of accretion and outflow.

\subsection{Interstellar Scattering}

Perhaps unfortunately, \sgr\ is heavily scattered by interstellar plasma at centimeter and longer wavelengths \citep{Lo98,Bow06,Lu11}. The combination of compact emission and heavy scatter-broadening at centimeter wavelengths has impeded understanding of the geometry of \sgr\ and the processes responsible for its emission. 

Scattering of radio waves in the interstellar plasma results from small-scale fluctuations in electron density.
Evidence suggests that the scatterers are part of a power-law spatial spectrum of density fluctuations \citep{Arm95}.
In other words, the difference of electron density, between two nearby points, has variance that increases as a power-law with the separation of the two points.
Scattering often displays the Kolmogorov scaling index of $\alpha=5/3$ expected for a cascade of Alfv\'{e}n-wave turbulence \citep{Gold95,Lith01}.
The cascade is initiated by driving forces at a large spatial scale, the ``outer scale''; and is terminated by dissipation at a minimum scale, the ``inner scale''.
The inner scale may be a few hundred km in the interstellar medium \citep{Spa90}.
The measured $\approx$2:1 anisotropy of the scatter-broadening of \sgr\ is typical for heavily-scattered lines of sight \citep{Des01}; 
it may indicate that density fluctuations responsible for scattering are aligned with a large-scale magnetic field \citep{Des94,Gold95}. 

A power-law spectrum of turbulence imprints its power-law index upon the scattered image.
For a scattered point source, it affects the scaling of angular broadening with wavelength,
and the distribution of flux density with radius \citep{Arm95}.
In accord with the fundamental principle of synthesis imaging via interferometry,
the interferometric visibility as a function of baseline length
is the Fourier-conjugate of this distribution,
so the visibility as a function of baseline length reflects the power-law,
with long baselines reflecting small-scale structures.
The averaged visibility is expected to be zero, for baselines long enough to resolve a smooth scattered image.

\subsection{Observations of Scattering of \sgr}

Very-long baseline interferometry (VLBI) observations of \sgr\ reveal a smooth, elliptical-Gaussian image indicative of strong scattering \citep{Kri98,Bow04,Shen05}. The image angular size $\theta$ scales with observing wavelength $\lambda$ as $\lambda^2$ over a wavelength range of centimeters to a meter \citep{Lo85,Jau89,Kri93,Yu94,Shen05,Bow06,Doe08,Lu11}. 
The smoothness and scaling are consistent with predictions for scattering by density fluctuations in the interstellar plasma, but differs from the scaling of $\theta\propto \lambda^{11/5}$
expected if the fluctuations follow a Kolmogorov spectrum.
At shorter wavelengths, the angular size departs from this power-law, as source structure becomes important \citep{Doe08,Lu11}.
By combining observations over a range of wavelengths, observers have inferred a size for the source after deconvolution of the scattering disk,
to obtain a model-dependent intrinsic size for \sgr\ as a function of wavelength.
This deconvolution leads to intrinsic dimensions at 1.3 cm wavelength of about 1 mas, depending upon assumptions about the scattering material
\citep{Lo98,Shen05,Bow06,Lu11}.

\begin{figure}
\includegraphics[width=0.48\textwidth]{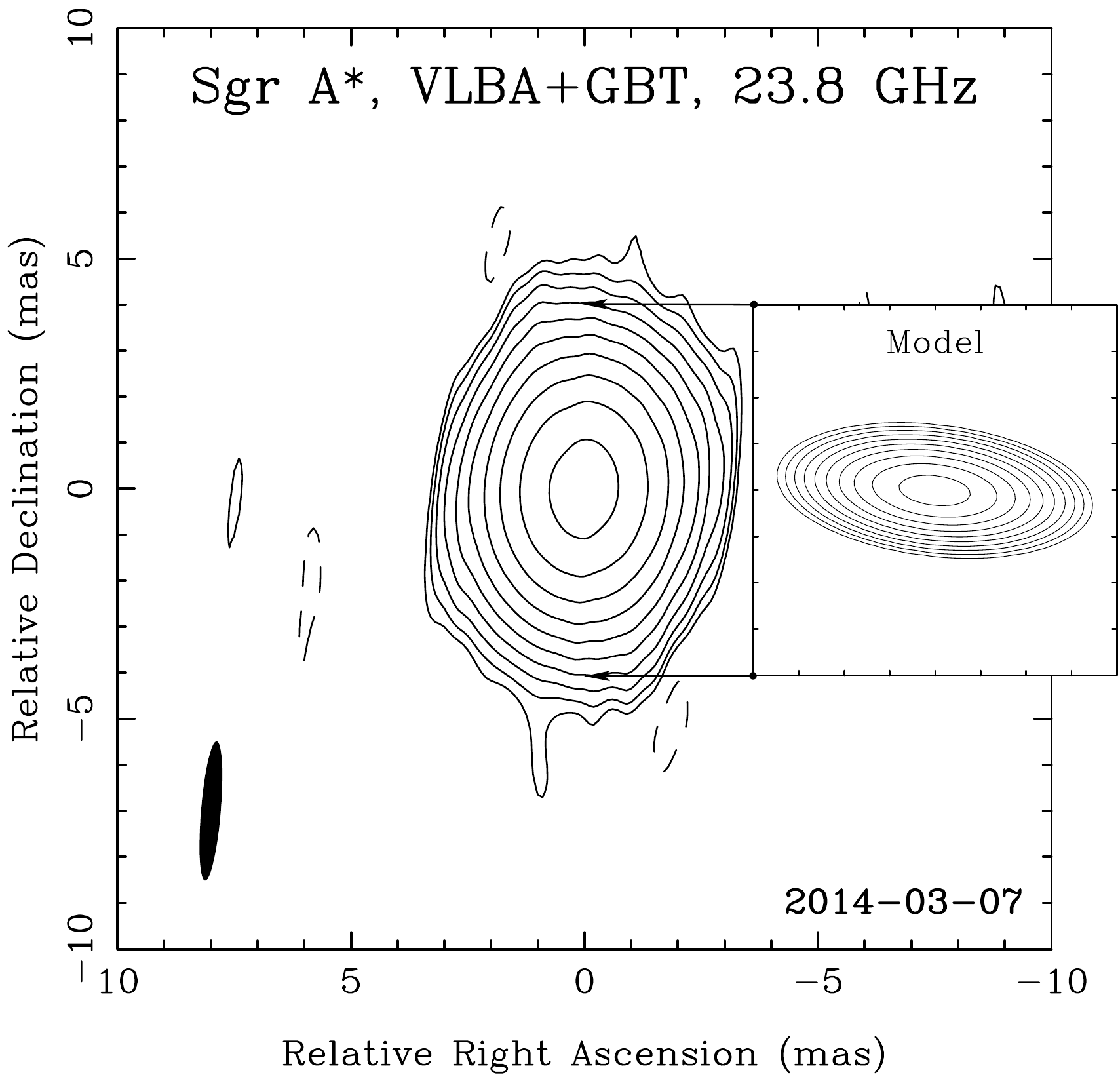}\\
\caption{Hybrid image of SgrA* from our naturally-weighted data at 1.3 cm, showing the scattering disk extended east-west. Substructure would appear as slight variations within the scattering disk. The contours of equal intensity are plotted starting from 0.14\% of the peak value of 190 mJy/beam with $\sqrt{2}$ steps. The restoring beam for the image, shown at lower left, is extended north-south because of the east-west extension of the array which provided detections; the beam has dimensions at half power level ($3.04\times0.42$) mas. Nearly all of the north-south extension of the image arises from our elliptical beam. The best-fitting model for the elliptical scattered image is shown as an inset.
\label{fig:image}}
\end{figure}

\subsection{Substructure in Scattering Disks}

Radio-wave scattering in the interstellar medium of \sgr\ is ``strong'': in other words, the multiple paths that the signal takes from source to observer differ in length by many wavelengths. 
Source images, or interferometric observations, that are subject to strong scattering
may be divided into 3 categories: snapshot, average, and ensemble-average regimes \citep{Nar89,Goo89}. 
In the ``snapshot'' regime, phase relationships among paths remain nearly constant during the observation, and speckles appear from interference. 
In the ``ensemble-average'' limit, an average over many of the possible
paths leads to a smooth, stable scattered image. The
``average'' or ``average-image'' regime lies between these two; in that regime, averaging has eliminated small-scale variations, but left large-scale variations intact. 
Averaging in time and frequency can shift an observation from the snapshot limit to the average-image regime, as can extended source structure.

For \sgr, observations at $\lambda=1.3\ {\rm}$ cm shorter than a few weeks are in the average-image regime because the source is extended. 
For a single VLBI observation spanning a few hours, substructure in the image should be nearly fixed.
Because such structures are smaller than the scattering disk, they modulate the scattered intensity, even on baselines long enough 
to resolve the average scattering disk.
This results in enhanced visibility on long baselines, with a random, noiselike character:
it averages out over times longer than that for Galactic rotation to carry the line of sight across the scattered image,
or a few weeks.
Consequently, theory predicts the root-mean-square visibility on long baselines,
and the average visibility on short baselines \citep{Nar89,Goo89,Joh13}.
Recent space VLBI observations using the Radioastron spacecraft \citep{Kard13}, on baseline projections up to 250,000\ km, show such substructure for the heavily-scattered Vela pulsar and PSR B0329+54, (C. R. Gwinn et al. in prep., M. V. Popov et al. in prep.). 
Moreover, \citet{Kel77} detected strong structure on scales smaller than the scattered image, in observations of \sgr\ at $\lambda=3.6\ {\rm cm}$.
Those results suggested the observations reported here.

\section{Observations}\label{sec:observations}

We observed \sgr\ using the Very Long Baseline Array (VLBA) in concert with the Green Bank Telescope (GBT) at 1.3 cm wavelength on 7 March 2014. For the observations we used the new NRAO Roach Digital Backend with a digital down-converter, and the Mark5C recorder at a bit rate of 2 Gbps. We observed in 4 contiguous 128-MHz channels with a central frequency of 23.8\ GHz, well separated from the H$_2$O line.  We recorded left circular polarization with a total bandwidth of 512 MHz, and 2-bit sampling. The recent improvements in the recorded bandwidth and the backend have at least doubled the sensitivity of the GBT and VLBA for VLBI observations. The total observing time on \sgr\ was about 3 hours. We also observed the compact extragalactic radio source 1730$-$130 as a calibration source, and obtained strong fringes for it on all baselines.

\begin{figure}
\includegraphics[width=0.48\textwidth]{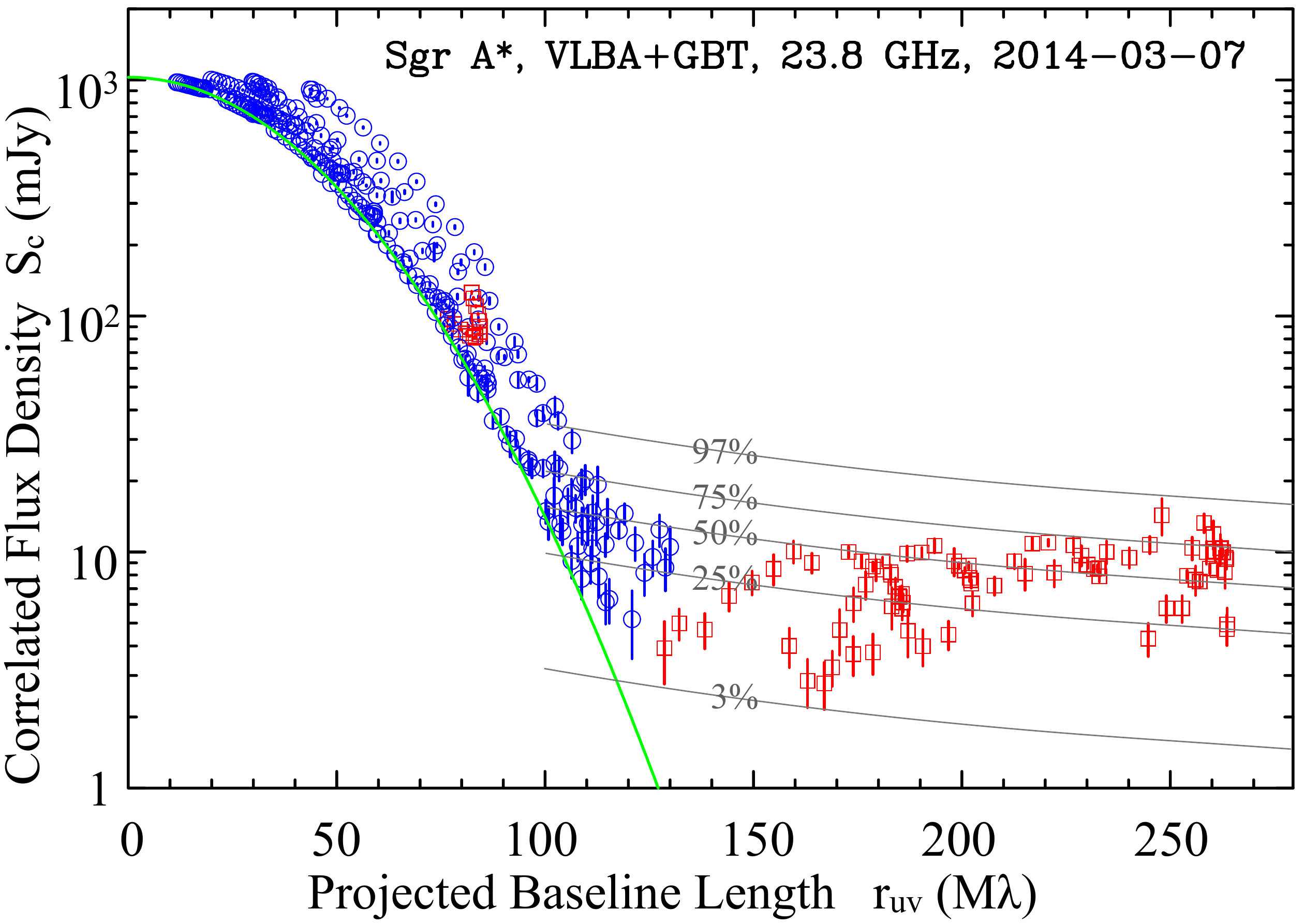}\\
\caption{
Correlated flux density of \sgr\ at $\lambda=1.3$\ cm plotted against baseline length. Squares show the sensitive GBT-VLBA baselines, circles VLBA-VLBA baselines. Each point represents a 15-min vector average for one baseline after self-calibration; error bars show statistical $\pm 1\sigma$. 
The green curve shows the correlated flux density of the average scattering disk for an East-West baseline, as described in Sec.\ \ref{sec:observations}. The gray curves show quantiles of the predicted distribution of correlated flux density from substructure, with source and scattering parameters from \citet{Bow06,Bow14b}, as described in Sec.\ \ref{sec:comparison} below.
\label{fig:radplot}}
\end{figure}

\begin{figure}[h]
\includegraphics[width=0.48\textwidth]{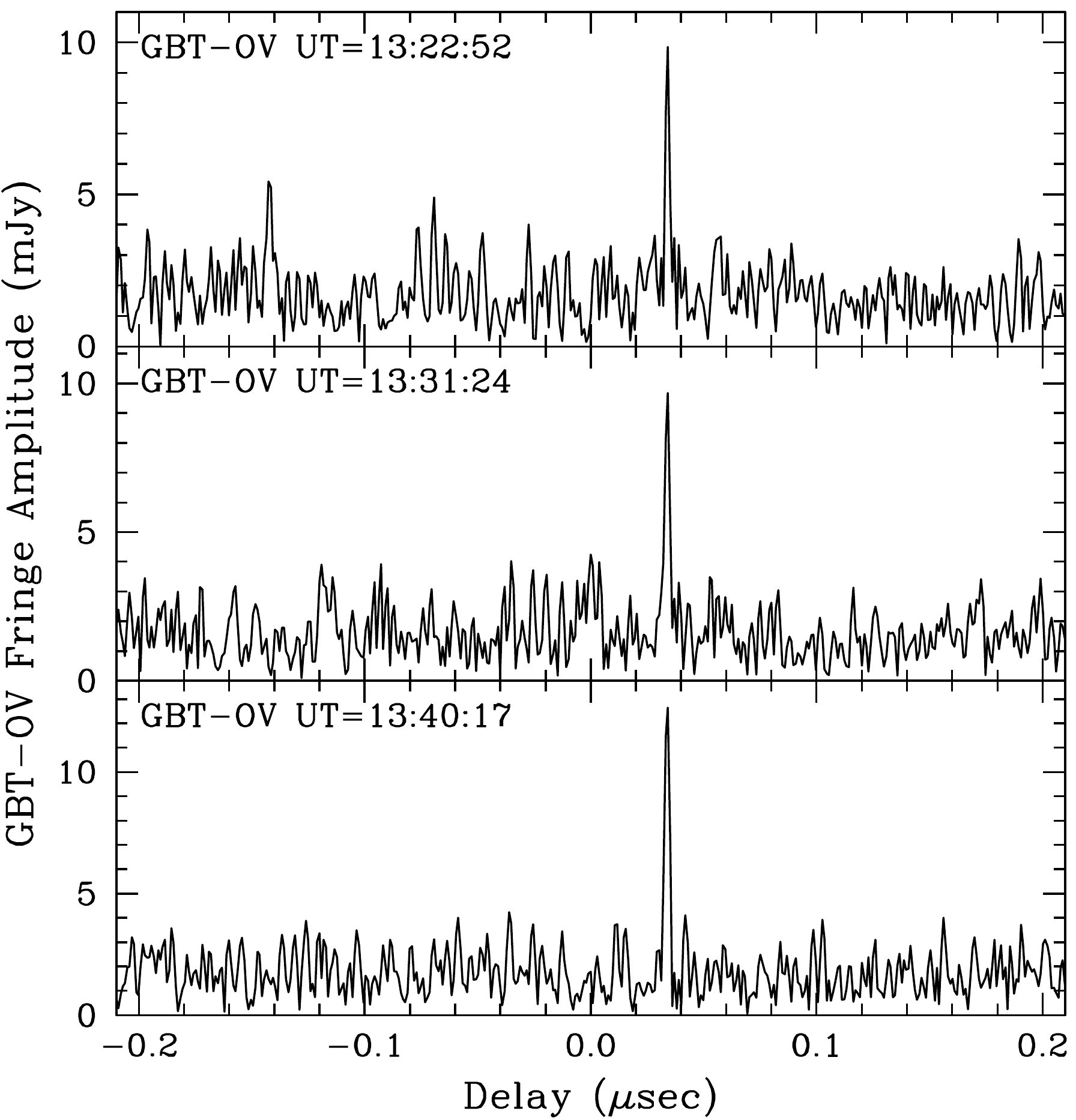}
\caption{Fringes on \sgr\ on the 3000-km GBT-Owens Valley baseline.
To make this plot, we Fourier transformed the data from the correlator, expressed as the complex visibility function as a function of frequency channel and time, to the domain of delay and fringe rate. We located the highest peaks in that domain for each interval, and display them here as a function of delay. 
The average scattering disk has correlated flux density, of ${<}\,10^{-6}$\ mJy on this baseline.
The probability of false detection is ${<}\,10^{-9}$ for each of the 3 peaks. 
}
\label{fig:fringes_ov}
\end{figure}

We performed conventional a priori calibration in AIPS including antenna-based fringe fitting \citep{Grei03}, and self-calibration and hybrid imaging in DIFMAP \citep{She97}.
Figure\ \ref{fig:image} shows the result.
We fitted the size of the scattering disk to our data in the visibility domain using DIFMAP, and found a size for the average scattered image of 2.26$\times$0.92 mas (full width at half maximum) with major axis at a position angle of 84 degrees, consistent with previous results at our observing wavelength \citep{Bow04,Shen05,Bow06,Lu11}.

Our long GBT-VLBA baselines completely resolve the average scattering disk, but nevertheless revealed a significant excess of correlated flux density (see Figure\ \ref{fig:radplot}). To verify that these detections are robust, we performed a careful baseline-based fringe-search of the data. Figure\ \ref{fig:fringes_ov} shows an example: peak correlated flux density for 3 consecutive 512-s intervals, for the GBT-Owens Valley baseline. The projected baseline is about 255\ M$\lambda$. The peaks range from 8.2 to 10.2 times the root-mean-square noise. The probability of attaining such high amplitudes by chance is less than $10^{-9}$ in any interval \citep{TMS,Pet11}. Moreover, we detect the peak in 3 consecutive fringing intervals, at the same fringe rate and the same delay, and near the values expected from geometric models. We detect the fringes independently in each of the 4 frequency channels. Results on other long GBT-VLBA baselines with detections were similar to these.

A fit in the visibility domain to a single elliptical Gaussian component, representing only the scattering disk, yielded reduced $\chi^2=1.91$, indicating an unsatisfactory fit. This model could not explain the data at projected spacings longer than 120\ M$\lambda$, as shown in Figure \ref{fig:radplot}. Inclusion of a second $\delta$-function component to the model with $\approx 10$\ mJy amplitude yielded reduced $\chi^2=1.28$. Using the method suggested by \citet{Kov05}, we found an upper limit to the size of this more compact component of about 0.3 mas, in the east-west direction. Thus, analysis by both fringe-search and model-fitting confirms the presence of highly compact substructure.

We do not believe that the observed fringes could arise from a background source, or an intervening source along the line of sight.
A background extragalactic source would be scattered as much as \sgr\ or more.
A foreground source would have to coincide with \sgr\ to the remarkable angular accuracy demanded by fringe rate and delay.
A pulsar with a flux density of 10\ mJy at $\lambda=1.3$\ cm would have been detected in previous surveys,
and an H$_2$O maser would be spectrally narrow.

\begin{figure}
\includegraphics[width=0.48\textwidth]{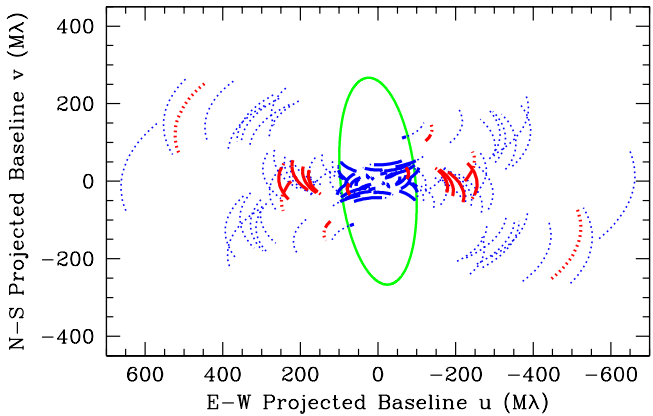}
\caption{Detections of fringes as a function of position in the plane of baselines perpendicular to the line of sight. GBT-VLBA baselines are shown as red lines, VLBA-VLBA as blue. Detections are solid lines, non-detections dotted. The green ellipse shows the best-fitting average scattering disk, expressed as the baseline where correlated flux density reaches $e^{-4}\approx0.018$ of maximum. Detections on shorter VLBA baselines, within the ellipse, show the large-scale scattering disk. Detections on long GBT-VLBA baselines indicate substructure within the disk.
}
\label{fig:uvdplot}
\end{figure}

We did not detect fringes on \sgr\ on the GBT-Mauna Kea baseline, at a 7-$\sigma$ upper limit of $\approx 5\ {\rm mJy}$. The statistics of the visibility were consistent with noise. We were not able to use data from
the GBT-Hancock baseline. Saturation or interference effects on this short baseline may have played a role. Sensitivity on baselines to northern antennas in the array, namely Brewster, Hancock, and the GBT, was significantly reduced due to the low declination of \sgr. 
Figure\ \ref{fig:uvdplot} shows all of our projected baselines, indicating detections.

\section{Comparison with Theoretical Models}\label{sec:comparison}

Although the time and frequency averaging of our data analysis would place us in the snapshot regime for \sgr, the source size puts us well into the average-image regime. The large-scale refractive variations left intact in this regime are presumably responsible for the observed substructure in the scattering disk. Such variations would be stochastic with a correlation length (in projected baseline) of approximately the diffractive scale (hundreds of km), they would be broadband, and they would persist over the refractive timescale (weeks). Thus, our current detections sample only a few independent elements of the substructure. 

The expected level of refractive substructure depends on the scattering geometry and anisotropy, the spectrum of density fluctuations in the scattering material, and the intrinsic source structure. Because the ensemble-average scattered image depends on these parameters in a different way, our additional measurements can break subtle parameter degeneracies and provide a deeper understanding of the scattering.

\begin{figure}
\includegraphics[width=0.48\textwidth]{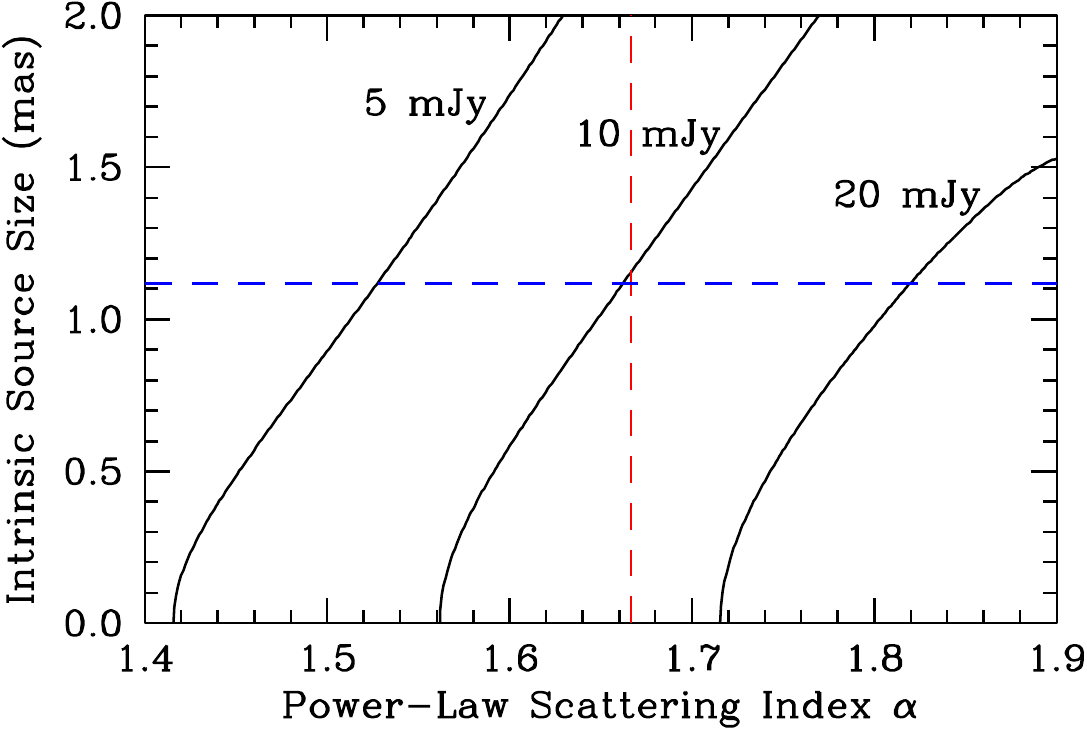}
  \caption{ 
Expected root-mean-square level of refractive noise, $S_{\rm rms}$, on a 3000-km E-W baseline ($230\ {\rm G}\lambda$) as a function of the power-law index of the  spectrum of density fluctuations, $\alpha$, and the intrinsic size (FWHM) of the source $\theta_{\rm src}$ along the baseline direction. 
Each of the three curves is calculated assuming that the diffractive scale and anisotropy are determined by extrapolating longer-wavelength measurements; the scattering geometry determined by \citet{Bow14a} is assumed. 
The horizontal dashed line shows the intrinsic size inferred from deconvolution \citep{Bow06}; the dashed vertical line shows the index $\alpha$ expected for a Kolmogorov spectrum. 
Samples on other baselines, or at another frequency, would break the degeneracy between $\alpha$ and $\theta_{\rm src}$.
Note that, because the amplitude of refractive noise is drawn from a Rayleigh distribution, its mean amplitude is $(\sqrt{\pi}/2) S_{\rm rms} \approx 0.89 S_{\rm rms}$ and its median amplitude is $\sqrt{\ln 2} S_{\rm rms}\approx 0.83 S_{\rm rms}$.
}
\label{fig:ThetaAlpha}
\end{figure}

Our present measurements set constraints on the spectrum of density fluctuations that scatters \sgr.
The observed scaling $\theta\propto \lambda^2$ of image size $\theta$ with wavelength $\lambda$ is consistent with either $\alpha=2$, or with any ``shallow'' spectrum ($\alpha<2$) and an inner scale larger than the diffractive scale: $r_{\rm in} > r_{\rm diff}=300\ {\rm km}$.  
Our detection of substructure indicates that the spectrum is shallow with an inner scale larger than the diffractive scale (see Figure\ \ref{fig:ThetaAlpha}). A large inner scale increases the level of refractive noise, so the inner scale cannot substantially exceed the diffractive scale. However, the effect of a large inner scale is slight, scaling the noise by $(r_{\rm in}/r_{\rm diff})^{1-\alpha/2}$. Given the paucity of independent samples in our observations, and the stochastic character of the signal, we can only tentatively conclude that $r_{\rm in} < 10^4\ {\rm km}$. However, additional data that determine the root-mean square flux density $S_{\rm rms}$ of the stochastic substructure to ${\sim}10\%$, and $\alpha$ from its scaling with baseline, could estimate the inner scale $r_{\rm in}$ to within a factor of 2.

Similarly, a small outer scale of the turbulence (relative to the ${\sim}3\ {\rm AU}$ refractive scale) would act to suppress the level of refractive noise \citep{Nar89,Goo89}. Hence, the lack of apparent suppression suggests an outer scale that is at least an AU, as expected from other refractive studies \citep{Arm95}. 

Finally, as with other scintillation effects, the substructure is also affected by the source size \citep{Joh13}. As a refractive effect, the long-baseline noise is quenched for a source that exceeds the refractive scale. The noise is approximately reduced by the squared ratio of the scattered size of a point source to the scattered size of the source. Importantly, this suppression factor is independent of baseline length and is only sensitive to source structure parallel to the baseline. Thus, as with source size estimates via deconvolution, our current measurements are most sensitive to source structure in the East-West direction.

Perhaps the simplest assumption for source and scattering is a point source ($\theta_{\rm src}=0$) scattered by a Kolmogorov spectrum of turbulence ($\alpha=5/3$). Under these assumptions, theory predicts substructure with root-mean square flux density of $S_{\rm rms}\approx 15$\ mJy on our 3000-km baselines. However, including the $1\ {\rm mas}$ source size estimated from deconvolution \citep{Bow06}, we expect $S_{\rm rms}\approx 10\ {\rm mJy}$, which is more compatible with our current measurements. 
This model predicts the distribution of correlated flux density shown by the gray curves in Figure\ \ref{fig:radplot}.
Figure\ \ref{fig:ThetaAlpha} shows the combinations of $\alpha$ and $\theta_{\rm src}$ that are consistent with this level of substructure, and for values of $S_{\rm rms}$ differing by factors of 2.

\section{Summary} 

We have detected substructure within the scattered image of \sgr\ at 1.3 cm wavelength, providing fresh insight into the scattering and structure of this supermassive black hole. Our estimates of source structure at 1.3 cm are complementary to those obtained by deconvolution of the ensemble-average scattering disk, because deconvolution must extrapolate the effects of scattering into ranges of wavelength where they cannot be measured directly. Moreover, our measurements indicate that the turbulent spectrum of the scattering material is shallow, so the effects of scattering at shorter wavelengths may be \emph{weaker} than previously supposed.  
We find that the inner scale of that turbulent spectrum is greater than 300~km, but less than $10^4$\ km. 
We find that the size of the source is consistent with the 1~mas estimates from deconvolution.
Additional measurements of substructure over a wider range of baselines and wavelengths can precisely determine the spectrum of density fluctuations,
and the intrinsic size of \sgr\  at centimeter wavelengths.
These will be of critical importance for efforts to image the black hole on event-horizon scales \citep{Doe08,Fi11,Fi14}.

\acknowledgments

We thank Leonid Petrov for assistance with detections and their significance, Mikhail Popov, Nikolai Kardashev, Ken Kellermann, Richard Porcas, and Mark Reid for essential discussions, and the referee for encouraging suggestions.
We thank the NRAO staff for supporting our observations in their usual highly professional and friendly manner.
The Robert C. Byrd Green Bank Telescope and the Very Long Baseline Array are operated by the National Radio Astronomy Observatory, which is a facility of the National Science Foundation, operated under cooperative agreement by Associated Universities Inc. C.R.G. acknowledges support of the US National Science Foundation (AST-1008865). Y.Y.K. was supported in part by the Dynasty Foundation.

Facilities: \facility{GBT}, \facility{VLBA}.

\end{document}